\documentclass[12pt]{iopart}
\usepackage{graphicx}
\usepackage[english]{babel}
\newcommand{\degree}{${^\circ}~$}
\begin{document}
\title{Unzipping graphene: Extendend defects by ion irradiation\\
}

\author{S. Akc\"oltekin$^1$, H. Bukowska$^1$, T. Peters$^1$, O. Osmani$^{1,2}$, I. Monnet$^3$, I. Alzaher$^3$, B. Ban d'Etat$^3$, H. Lebius$^3$ and M. Schleberger$^1$ \footnote{electronic address: marika.schleberger@uni-due.de}} 
\address{$^1$Fakult\"at f\"ur Physik and CeNIDE, Universit\"at Duisburg-Essen, D-47048 Duisburg, Germany}
\address{$^2$ Technical University of Kaiserslautern, OPTIMAS Research Center, D-67653 Kaiserslautern, Germany}
\address{$^3$CIMAP (CEA-CNRS-ENSICAEN-UCBN), 14070 Caen Cedex 5, France}

{\bf
Many of the proposed future applications of graphene require the controlled introduction of defects into its perfect lattice \cite{Geim2009,NewsViews}. Energetic ions provide one way of achieving this challenging goal. Single heavy ions with kinetic energies in the 100 MeV range will produce nanometer-sized defects on dielectric but generally not on crystalline metal surfaces \cite{Toulemonde1992,Akcoeltekin2007}. In a metal the ion-induced electronic excitations are efficiently dissipated by the conduction electrons before the transfer of energy to the lattice atoms sets in. Therefore, graphene is not expected to be irradiation sensitive beyond the creation of point defects. Here we show that graphene on a dielectric substrate sustains major modifications if irradiated under oblique angles. Due to a combination of defect creation in the graphene layer and hillock creation in the substrate, graphene is split and folded along the ion track yielding double layer nanoribbons. Our results indicate that the radiation hardness of graphene devices is questionable but also open up a new way of introducing extended low-dimensional defects in a controlled way.}

Experimental data on ion induced modifications of graphene is still surprisingly scarce \cite{Tapaszto2008,Stolyarova2009}
but there has been much work on ion induced defects in other carbon based nanomaterials such as fullerenes and carbon nanotubes \cite{Banhart1999,Arkady2007,Krasheninnikov2010}. The carbon network readily reconstructs around any ion induced defect and bandstructure as well as material properties will change accordingly. In graphene this would be extremly useful with respect to tailoring its electronic properties for spintronics \cite{Okada2006} or ultrafast transistors \cite{Liao2010}, but even direct guiding of charges may become feasible if extended defects can be realized \cite{Batzill2010}. 

Depending on the kinetic energy of the projectile, there are two different mechanisms to create defects: In the low energy regime, defects are mainly introduced by collisions of the ion with the carbon atoms ({\it nuclear stopping}), {\it i.e.}~a direct hit. In the high energy regime electronic excitations play the dominant role ({\it electronic stopping}). In this process much more energy can be transferred and direct hits are no longer required. At $E_{kin}$=100 MeV, a Xe ion deposits more than 99\% of its energy into the solid by electronic stopping \cite{SRIM}. In a dielectric material like SiO$_2$ or SrTiO$_3$ this electronic excitation is eventually transferred to the lattice and the strong localized energy deposition leads to a molten or even vaporized region surrounding a swift heavy ion's (SHI) trajectory, the so-called ion track. This is followed by a rapid quenching due to the large temperature difference between the track and the surrounding material \cite{Kluth2008}. 

Under perpendicular incidence the typical remnants of this process are nanometersized hillocks (height depending on the material, in SrTiO${_3}~ h \simeq 4 \pm 1 $ nm, radius 5 nm $\leq r \leq$ 15 nm) at the impact point of each ion, see \fref{figure1} a. In the case of glancing angle irradiation a part of the ion track appears on the surface \cite{Akcoeltekin2007}. It consists of separate nanometer-sized hillocks with similar dimensions (in SrTiO${_3}~ h \simeq 3 \pm 1 $ nm). The length of the chain can be controlled by the angle of incidence \cite{Akcoeltekin2008} and can reach a micron or more. The peculiar periodicity of the hillocks is attributed to the spatial electronic density distribution \cite{Akcoeltekin2007} but will not be discussed here any further as it does not seem to play a role for the folding process.

We studied the effects of SHI irradiation of graphene exfoliated on SrTiO$_3$ \cite{SAkcoeltekin2009} by means of atomic force microscopy (AFM). If irradiated under $\Theta=90$\degree the typical hillocks that appear on this substrate can also be found - now {\em underneath} the graphene (see \fref{figure2}a). The hillocks on the substrate are 3.2$\pm$0.9 nm, the ones under the graphene are 2.7$\pm$0.7 nm in height. Apart from the straining no damage of the graphene is observed at this resolution. If the angle of irradiation is changed to $\Theta=1$\degree, characteristic extended damage patterns occur (see \fref{figure2}b). As the bare substrate is clearly visible in the middle (see also supporting material) the damage obviously consists of folded graphene. The orientation of the folded part coincides with the direction of the ion beam. The dimensions of the folded graphene are typically several 100 nm in length and about 50 nm in width. Multilayer graphene sheets are not folded as frequently as monolayers. As an example, in \fref{figure2}c a graphene trilayer hit by two single ions is shown. One ion caused the typical folding pattern while the other did not produce any visible folding damage. 

To explain the striking difference between perpendicular and glancing angle irradiations let us first discuss if a single hillock could break graphene. The breaking strength of defect free graphene (intrinsic strength) has been determined by means of indentation experiments with an atomic force microscope to be $\sigma^{2D}_{Max}$ = 55 N/m \cite{CLee2008} corresponding to the Young modulus of graphene $E \simeq 1$ TPa \cite{Booth2008}. In our case the indentation is negligible and not sufficient to cause any damage. From the glancing angle data we can directly determine an upper limit for the strain of the graphene layer induced by the chain of nano hillocks. We apply the linear model where $\epsilon=\sigma/E=\Delta l/l_0$. The exposed region (see \fref{figure2}a) is a direct measure for the original length $l_0 \simeq$ 60 nm of the unstrained graphene layer. The strained graphene covered a series of hillocks with a height of $h \leq $4 nm and was thus strained by $\Delta l \leq $ 4.6 nm assuming that the strain is equally distributed over the whole layer. We find $\epsilon\leq 7.6\%$ which is clearly below the typical inelastic limits of 10\% - 14\% for graphitic nano-materials \cite{Walters1999,Yu2000}.

However, the material of the hillocks may reach the boiling temperature and would then exert a pressure on the graphene. The single hillocks have typical dimensions comparable to those of an AFM tip and the geometry corresponds to a circular membrane clamped around the entire circumference. We can therefore adapt the approach proposed by Lee {\it et al.} to estimate the maximum stress on top of the hillock as a function of the pressure:
$$\sigma^{2D} = \sqrt{\frac{p~r~E^{2D}}{2}} \simeq 29~\rm{N/m}$$
The value for $E^{2D}$ has been obtained from the $3D$ value by multiplication with the nominal thickness of a single graphene sheet of 3.35 \AA, $r$ is the radius of the hillock and the pressure in an ideal gas approximation is $p = \rho RT =$ 1.25 GPa with $T_{boil}^{SrO}$ = 3300 K and $R = 80$ J/(K$\cdot$kg)$^{-1}$. Note that $\sigma^{2D}$ is significantly lower than the maximum stress for free-standing monolayer graphene.


Another possible reason for the folding could be some direct interplay between the ion induced electronic excitation and the pushing hillock.
This is however very unlikely, as there is a substantial time gap between any electronic processes in either the graphene or the substrate (on the order of a few fs) and the evolution of the hillocks which takes place only after some 10 ps \cite{Medvedev2010}. Neither the irradiation of graphene with 0.7 MeV protons yielded any folding \cite{Stolyarova2009} nor glancing angle experiments with SHI on graphite \cite{Liu2002,Akcoeltekin2009}. From this and the discussion above we conclude the following: A hillock alone cannot overcome the intrinsic strength of a graphene monolayer, but the folding is not a direct result of the interaction of the ion with the graphene either.

To further elucidate the mechanism we discuss the peculiarities of glancing angle irradiations next. At a glancing angle of $\Theta=1$\degree$\pm~0.5$\degree the ion will pass through at least 30 carbon bonds. Their high charge density offers an energy loss channel for the projectile via electronic excitations such as plasmon modes. Theoretical calculations have shown that the 
transferred energies easily reach 50 eV in the case of protons hitting a carbon bond \cite{Krasheninnikov2007}. For our case of SHI the transferred energy will be even higher. The minimum energy required to displace an atom in the $sp^2$ bonded graphene is 15-20 eV \cite{Banhart1999} 
and the typical energy for the formation of a stable defect like a reconstruction around a single vacancy or
a pentagon/heptagon reconstruction is around 5-7 eV \cite{Lusk2008}. Therefore, the passing ion will give rise to the creation of multiple {\it aligned} defects and the subsequent reconstruction of the carbon network along its track (see \fref{figure3}a). 

The changes in elastic properties of graphene under swift heavy ion irradiation have not yet been studied but 
defect-rich carbon materials such as nanotubes are known to deteriorate. The tensile strength of defective carbon nano tubes can be reduced by a factor of two \cite{Sammalkorpi2004,Zhang2005}. The chain of hillocks coincides exactly with the line of extended defects and would then indeed be able to push through the graphene and unzip it along the track as visualized in \fref{figure3}b. This is in agreement with our finding that a bi- or trilayer is less frequently ripped apart as the required force simply scales with the number of layers.

In conclusion we have shown that SHI irradiation of exfoliated graphene under perpendicular incidence does not result in any extended damage. After irradiation under glancing angles extended characteristic damage patterns are created. The damage could be clearly identified as regions of backfolded graphene. We propose that the folding becomes possible because the extendend defects created in the graphene by electronic excitations overlap spatially with the hillock chain created in the substrate at a later time. The unique properties of swift heavy ions with respect to range and energy loss mechanism thus make the controlled introduction of extended defects in graphene possible.

By using a metallic substrate which does not form hillocks under irradiation or by omitting the substrate altogether it might be possible to create these extended defects without breaking the graphene. Future irradiation experiments on {\it e.g.} graphene on metal single crystal substrates or on freestanding graphene \cite{Booth2008,Shivaraman2009} will reveal the exact nature of these extended defects, which are predicted to modify the electronic properties of graphene and have just recently been experimentally realised for the first time \cite{Batzill2010}. 

Finally, we would like to point out that our results imply that any graphene based devices which use a dielectric as a substrate will not be stable against ionizing particle radiation. This is in contrast to the recently stated stability against photon irradiation \cite{Li2008} and must be taken into account in {\it e.g.} future space applications.\\

{\small
\noindent
Methods\\
\noindent
Sample preparation\\
\noindent
For this experiment we have prepared several graphene crystallites on SrTiO$_3$(100)(Crystec,
Berlin) by mechanical exfoliation with adhesive tape from a graphite crystal (HOPG, Mateck).
An optical microscope was used to identify regions with graphene flakes, which is possible also
on SrTiO$_3$ despite the lower optical contrast \cite{SAkcoeltekin2009}. The samples were then transferred to a
Raman spectrometer (LabRam Jobin Yvon/Horiba). We used radiation with a wavelength
of 633 nm ($E_L=1,961$ eV) from a HeNe laser, a 100x objective and avoided excessive heating
of the graphene by limiting the laser power to $P_{max} = 4$ mW. Raman spectra were taken
from all samples under ambient conditions with a grating of 1800 grooves per mm yielding a
spectral resolution of $\simeq$ 2 cm$^{-1}$. The spot size was $\simeq$ 1 $\mu$m$^2$. As references we used substrates
without graphene as well as the corresponding spectra from HOPG. For calibration
purposes a Si single crystal was used. Raman spectra before irradiation showed the SLG fingerprint peaks at 2750 cm$^{-1}$ and no 
$D$ peak at 1350 cm$^{-1}$ indicating defect free monolayers of graphene.\\
\noindent
Irradiation\\
\noindent
The irradiation of the samples took place at the GANIL facility, Caen (France) at the beamline IRRSUD 
with Xe$^{23+}$ and Pb$^{29+}$ Ions at a total kinetic energy of 91 MeV and 104 MeV, respectively. This corresponds to a stopping power of $dE/dx = 21 - 23$ keV nm${-1}$ \cite{SRIM}. Angles of incidence were $\Theta = 90$\degree and $\Theta = 1$\degree$\pm~0.3$\degree with respect to the surface. Typical fluences were on the order of 1$\times 10^9$ ions per cm$^2$ ensuring that defects can be attributed to single impact events but are still numerous enough for statistical analysis.
\\
\noindent
Experimental\\
\noindent
An UHV-AFM (Omicron) was used to characterize the samples before irradiation. After irradiation an ambient AFM (VEECO-Dimension 3100) was used. 
Images were recorded in non-contact and tapping mode, respectively. We used standard Si tips (NCHR, Nanosensors) and typical operating parameters were: $df$= -10 Hz; drive amplitude 100 mV, amplitude set point 1 - 1.2 V, scan rate 1 Hz. Image processing was done by WSxM \cite{WSxM}; all images shown are raw data except for plane subtraction.}


\section*{Acknowledgement}
We thank the SFB 616: {\it Energy dissipation at surfaces} for financial support. The experiments were performed at the IRRSUD beamline of the Grand Accelerateur National d'Ions Lourds (GANIL), Caen, France. We thank K.R. Dey for her support at the beamline and A. Reichert for providing  \fref{figure3}. Correspondance and requests for materials should be addressed to M.S.

\section*{Author contributions}
S.A., H.L. and M.S. designed the experiment, which was conceived by M.S. The graphene samples were prepared by S.A. The irradiations were performed by all authors except for H.B. and O.O.; H.B. and S.A. recorded the AFM images. M.S., H.B., O.O. and S.A. analyzed the data and M.S. wrote the paper. All authors discussed the results and implications and commented on the manuscript.

\section*{Additional information}
The authors declare no competing financial interest.

\newpage
{\sffamily
\begin{figure}[ht!]
    \centering
    \includegraphics[width=0.7\textwidth]{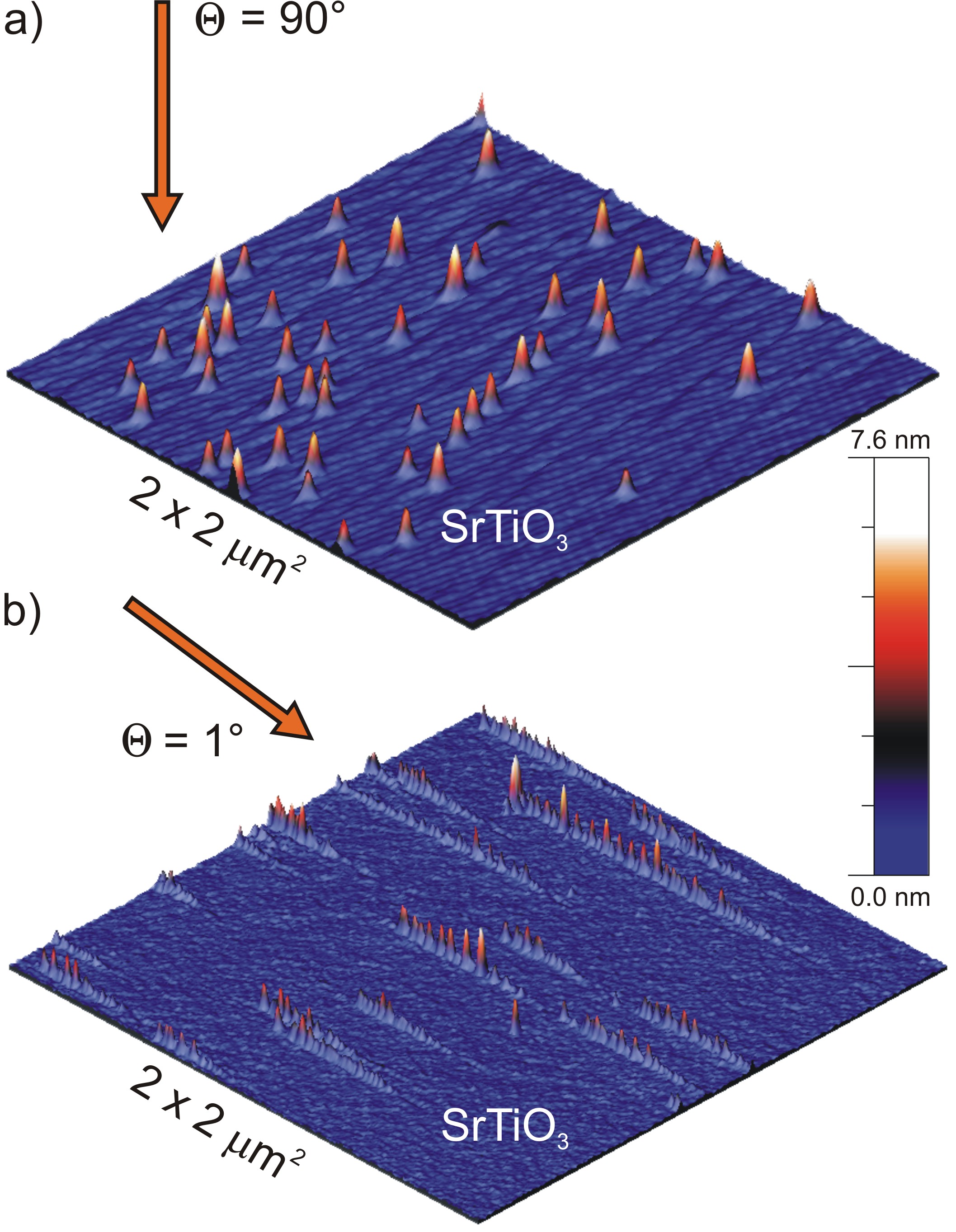}
    \caption{3D-visualization from non-contact AFM images of swift heavy-ion induced nano hillocks on SrTiO$_3$. The $z$-scale is enhanced with respect to $x$ and $y$ to show ion-induced surface modifications. {\bf a}, The irradiation took place under an angle of $\Theta=90$\degree as indicated by the arrow. Each hillock is produced by an individual ion. {\bf b}, Here, the irradiation took place under a glancing angle of $\Theta=1$\degree and each of the chains is produced by a single ion.}
    \label{figure1}
\end{figure}

\begin{figure}[ht!]
    \centering
    \includegraphics[width=0.54\textwidth]{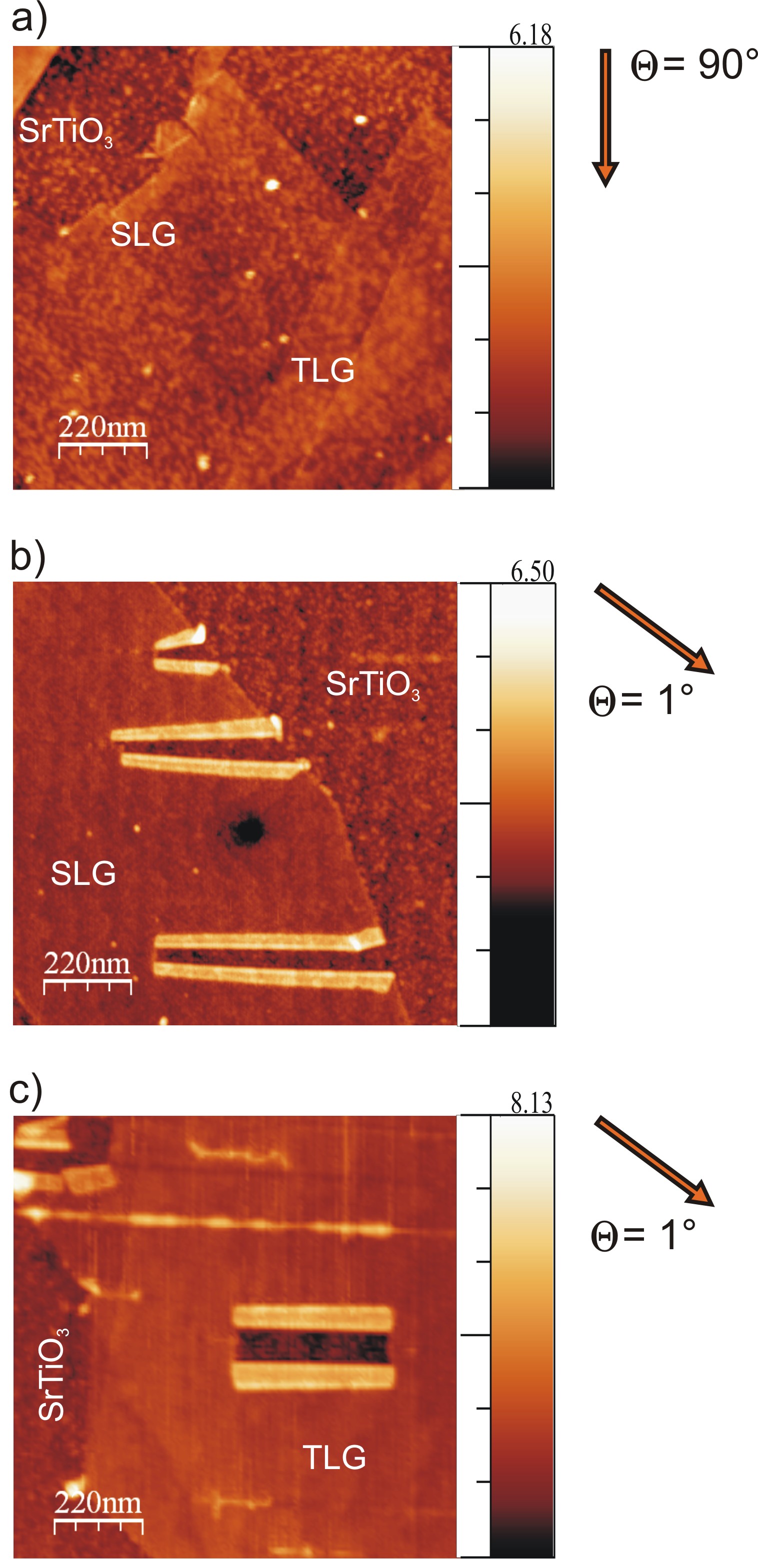}
   	  \caption{AFM images of SHI irradiated graphene on SrTiO$_3$, $z$-scale in nm. {\bf a-c},  In {\bf a} the sample was irradiated with Pb$^{29+}$ ions at 103 MeV under an angle of $\Theta=90$\degree with respect to the surface: hillocks are created but the graphene exhibits no folding damage. In {\bf b} and {\bf c} the sample was irradiated with Xe$^{23+}$ ions at 91 MeV under a glancing angle of $\Theta=1$\degree; ions were coming from the left. The graphene shows extendend damage in the form of folded parts. On single layer graphene (SLG) each ion track leads to folding. On multilayer graphene (here a trilayer marked with TLG) the efficiency is reduced.}
    \label{figure2}
\end{figure}

\begin{figure}[ht!]
    \centering
    \includegraphics[width=0.6\textwidth]{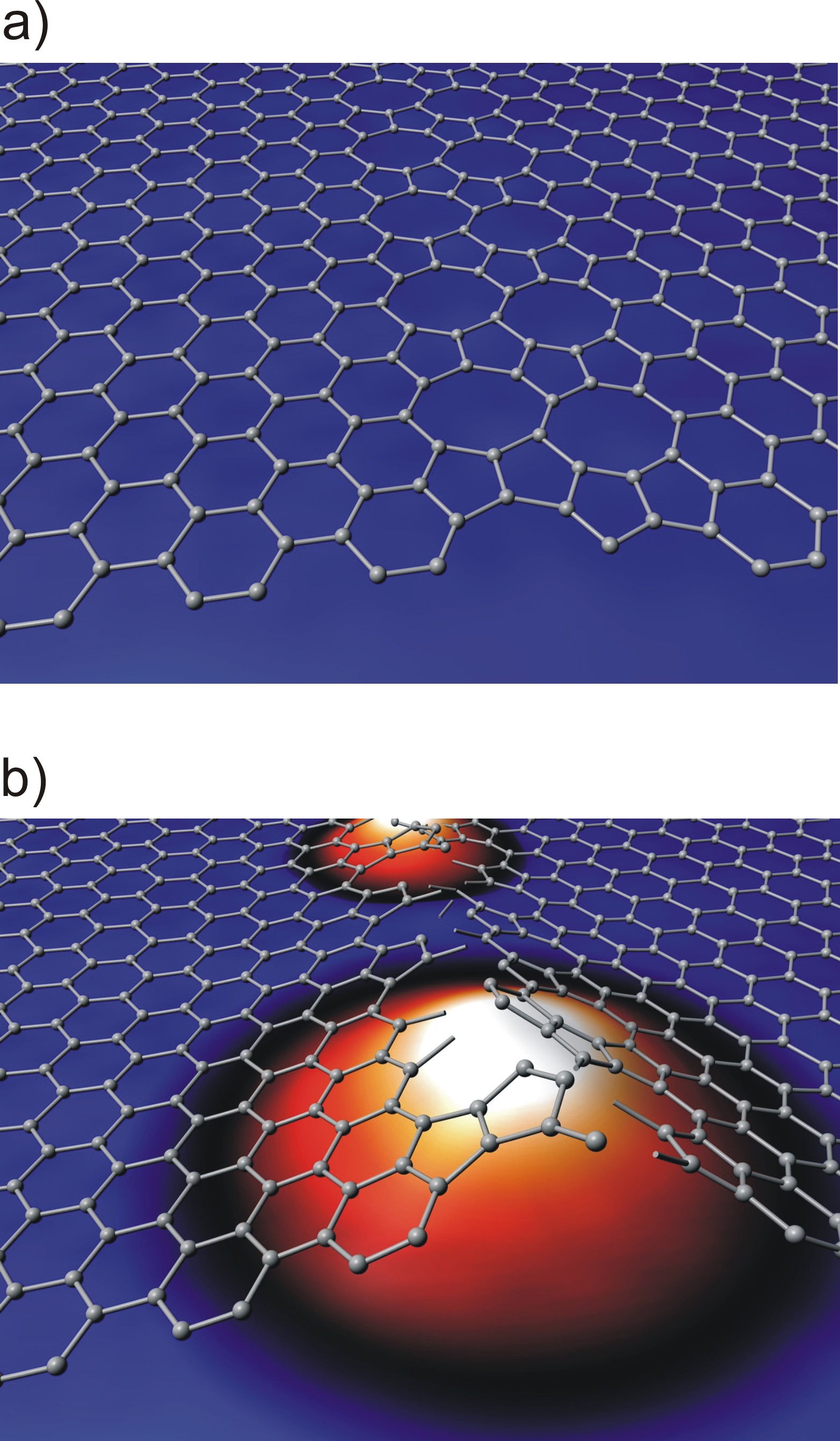}
	  \caption{{\bf a}, Possible extended defects in graphene which can be created by SHI irradiation under glancing angles. With a SHI, up to several 100 eV can be transferred from the projectile to each unit cell. The effective radius of the electronic excitations is on the order of some nm. {\bf b}, The hillock chain is created at much later times but overlaps spatially with the line-defect; the graphene can thus be zipped open along the ion track.}
    \label{figure3}
\end{figure}}

\end{document}